\documentclass[preprint2]{aastex}

\shorttitle{Gliese 581d is in the habitable zone}
\shortauthors{Wordsworth et al.}

\begin{document}


\title{Gliese 581d is the first discovered terrestrial-mass exoplanet in the habitable zone}


\author{Robin D. Wordsworth,$^{1\ast}$ Fran\c cois Forget,$^{1}$ Franck Selsis,$^{2,3}$\\
Ehouarn Millour,$^{1}$ Benjamin Charnay,$^{1}$ Jean-Baptiste Madeleine$^{1}$\\ \vspace{0.1in}
\normalsize{$^{1}$Laboratoire de M\'et\'eorologie Dynamique,}\\
\normalsize{Institut Pierre Simon Laplace, Paris, France}\\
\normalsize{$^{2}$CNRS, UMR 5804, Laboratoire d'Astrophysique de Bordeaux,}\\
\normalsize{2 rue de l'Observatoire, BP 89, F-33271 Floirac Cedex, France}\\
\normalsize{$^{3}$Universit\'e de Bordeaux, Observatoire Aquitain des Sciences de}\\
\normalsize{l'Univers, 2 rue de l'Observatoire, BP 89, F-33271 Floirac Cedex, France}\\
}

%\author{S. Djorgovski\altaffilmark{1,2,3} and Ivan R. King\altaffilmark{1}}
%\affil{Astronomy Department, University of California,
%    Berkeley, CA 94720}
%\author{C. D. Biemesderfer\altaffilmark{4,5}}
%\affil{National Optical Astronomy Observatories, Tucson, AZ 85719}
%\email{aastex-help@aas.org}
%\and
%\author{R. J. Hanisch\altaffilmark{5}}
%\affil{Space Telescope Science Institute, Baltimore, MD 21218}


%\altaffiltext{1}{Visiting Astronomer, Cerro Tololo Inter-American Observatory.
%CTIO is operated by AURA, Inc.\ under contract to the National Science
%\altaffiltext{2}{Society of Fellows, Harvard University.}
%\altaffiltext{3}{present address: Center for Astrophysics,
%    60 Garden Street, Cambridge, MA 02138}
%\altaffiltext{4}{Visiting Programmer, Space Telescope Science Institute}
%\altaffiltext{5}{Patron, Alonso's Bar and Grill}


\begin{abstract}
It has been suggested that the recently discovered exoplanet GJ581d might be able to support liquid water due to its relatively low mass and orbital distance. However, GJ581d receives 35\% less stellar energy than Mars and is probably locked in tidal resonance, with extremely low insolation at the poles and possibly a  permanent night side.  Under such conditions, it is unknown whether any habitable climate on the planet would be able to withstand global glaciation and / or atmospheric collapse.  Here we present three-dimensional climate simulations that demonstrate GJ581d will have a stable atmosphere and surface liquid water for a wide range of plausible cases, making it the first confirmed super-Earth (exoplanet of 2-10 Earth masses) in the habitable zone.  We find that atmospheres with over 10 bar CO$_2$ and varying amounts of background gas (e.g., N$_2$) yield global mean temperatures above 0~$^\circ$C for both land and ocean-covered surfaces. Based on the emitted IR radiation calculated by the model, we propose observational tests that will allow these cases to be distinguished from other possible scenarios in the future.
\end{abstract}



\maketitle

\section{Introduction}

The local red dwarf Gliese 581 (20.3 ly from the Sun, $M=0.31 M_{Sun}$, $L=0.0135 L_{Sun}$, spectral type M3V)  \citep{Hawley1997} has received intense interest over the last decade due to the low mass exoplanets discovered around it.
As of early 2011 it has been reported to host up to six planets \citep{Udry2007,Mayor2009,Vogt2010}.
One of these, GJ581g, was announced in September 2010 and estimated to be in the habitable zone (the orbital range in which a planet's atmosphere can warm the surface sufficiently to allow surface liquid water) \citep{Kasting1993,Pierrehumbert2011}. However, its discovery has been strongly disputed by other researchers, including the team responsible for finding the other four planets in the system \citep{Kerr2010,Tuomi2011}. For the moment, therefore, GJ581g remains unconfirmed.

GJ581d, in constrast, which was first discovered in 2007 and has a minimum mass between 5.6 and 7.1 $M_{Earth}$, has now been robustly confirmed by radial velocity (RV) observations \citep{Udry2007,Mayor2009,Vogt2010}.  Due to its greater distance from the host star, GJ581d was initially regarded as unlikely to have surface liquid water unless strong warming mechanisms due to e.g., CO$_2$ clouds \citep{Forget1997,Selsis2007} were present in its atmosphere.
Recently, simple one-dimensional radiative-convective studies \citep{Wordsworth2010b,vonParis2010,Kaltenegger2011} have suggested that a dense atmosphere could provide a significant greenhouse effect on GJ581d. However, the planet's tidal evolution poses a key problem for its habitability.

As it is most likely either in a pseudo-synchronous state with a rotation period that is a function of the eccentricity, or in  spin-orbit resonance like Mercury in our Solar System \citep{Leconte2010,Heller2011}, GJ581d should have extremely low insolation at its poles and possibly a permanent night side. Regions of low or zero insolation on a planet can act as cold traps where volatiles such as H$_2$O and CO$_2$ freeze out on the surface. A few previous studies \citep{Joshi1997,Joshi2003} have examined atmospheric collapse in 3D with simplified radiative transfer, but only for Earth-like atmospheric pressures or lower (0.1 to 1.5 bar). For low values of stellar insolation and large planetary radii, even dense CO$_2$ atmospheres will be prone to collapse, which could rule out a stable water cycle altogether for a super-Earth like GJ581d. To conclusively evaluate whether GJ581d is in the habitable zone, therefore, three-dimensional simulations using accurate radiative transfer are necessary.

Here we present global climate model (GCM) simulations we performed to assess this issue. In Section \ref{sec:method} we describe the model used. In Section \ref{sec:results} we describe our results, and in Section \ref{sec:conc} we discuss implications and propose future observational tests for the simulated habitable scenarios.

\section{Method} \label{sec:method}

In our simulations we made the initial hypothesis that GJ581d has a climate dominated by the greenhouse effects of CO$_2$ and / or H$_2$O, as is the case for all rocky planets with atmospheres in the Solar System (Venus, Earth and Mars). To assess the influence of water on the climate independently, we considered two classes of initial condition: a rocky planet with no water, and an ocean planet, where the surface is treated as an infinite water source.  CO$_2$ was taken as the primary constituent of the atmosphere and H$_2$O was allowed to vary freely, with surface ice / liquid and cloud formation (including radiative effects) taken into account for either gas when necessary. Restricting the composition of the atmosphere to two species in this way allows us to determine conservative conditions for habitability, as it neglects the warming due to other greenhouse gases like CH$_4$ or buffer gases like N$_2$ or Ar \citep{vonParis2010,Goldblatt2009,Li2009}.

The simulations were performed using a new type of GCM that we developed specifically for exoplanet and paleoclimate studies. It uses radiative transfer data generated directly from high resolution spectra, which allows the accurate simulation of climates for essentially any atmospheric cocktail of gases, aerosols and clouds for which optical data exists.
The dynamical core of the model was adapted from the LMD Mars GCM, which uses an enstrophy and total angular momentum conserving finite difference scheme \citep{Sadourny1975,Forget1999}. Scale-selective hyperdiffusion was used in the horizontal plane for stability. The planetary boundary layer was parameterised using the method of \citet{Mellor1982}  to calculate turbulent mixing, with the latent heat of H$_2$O also taken into account in the surface temperature calculations when necessary. A standard roughness coefficient of $z_0=1\times10^{-2}$ m was used for both rocky and ocean surfaces for simplicity, although we verified that our results were insensitive to variations in this parameter. Spatial resolution of $32\times32\times20$ in longitude, latitude and altitude was used for all simulations; we performed one comparison test at the highest rotation rate with $64\times64\times20$ resolution and found that the differences were small.

Our radiative transfer scheme was similar to that we developed previously for one-dimensional simulations \citep{Wordsworth2010,Wordsworth2010b}. For a given mixture of atmospheric gases, we computed high resolution spectra over a range of temperatures, pressures and gas mixing ratios. For this study we used a 6 $\times$ 9 $\times$ 7 temperature, pressure and H$_2$O volume mixing ratio grid with values $T = \{100, 150,\ldots, 350 \}$ K, $p  = \{ 10^{-3}, 10^{-2}, \ldots, 10^5 \} $ mbar and $q_{H_2O}=\{10^{-7}, 10^{-6}, \ldots, 10^{-1} \}$, respectively. The correlated-$k$ method was used to produce a smaller database of coefficients suitable for fast calculation in a GCM. The model used 38 spectral bands in the longwave and 36 in the shortwave, and sixteen points for the $g$-space integration, where $g$ is the cumulated distribution function of the absorption data for each band. In most simulations CO$_2$ was assumed to be the main constituent of the atmosphere, except the locally habitable ice planet experiments (see Section \ref{sec:results}), where N$_2$ was used. CO$_2$ collision-induced absorption was included using a parameterisation based on the most recent theoretical and experimental studies \citep{Wordsworth2010,Gruszka1998,Baranov2004}. For the stellar spectrum, we used the Virtual Planet Laboratory AD Leo data \citep{Segura2005}. AD Leo is a M-class star with effective temperature $T_{eff} = 3400$~K, which is acceptably close to the most recent estimates of $T_{eff} = 3498 \pm 56$~K \citep{vonBraun2011} for the purposes of climate modelling. A two-stream scheme \citep{toon1989} was used to account for the radiative effects of both clouds and Rayleigh scattering, as in \citet{Wordsworth2010b}.

In the water cycle and cloud modelling, care was taken to ensure that the parameterisations used were based on physical principles and not tuned to Earth-specific conditions. When this was not possible (as for e.g., the density of condensable cloud nuclei in the atmosphere $N_c$), we tested the sensitivity of our results to variations in those parameters.
Three tracer species were used in our simulations: CO$_2$ ice, H$_2$O ice and H$_2$O vapour. Tracers were freely advected in the atmosphere, subject to changes due to sublimation / evaporation and condensation and interaction with the surface. For both gases, condensation was assumed to occur when the atmospheric temperature dropped below the saturation temperature. Local mean CO$_2$ and H$_2$O cloud particle sizes were determined from the amount of condensed material and the density of condensable nuclei $N_{c}$. This parameter was set to $10^5$ kg$^{-1}$ in most of our simulations; we tested the effect of varying it over the range $10^4$ to $10^6$ kg$^{-1}$ and found that the maximum difference in mean surface temperature after 60 orbits was less than 5 K. As a further test of the robustness of our results, we also performed some tests with cloud radiative effects removed altogether (see Section 3).

Ice particles of both species were sedimented according to Stokes law \citep{Forget1999}. Below the stratosphere, adjustment was used to relax temperatures due to convection and / or condensation of CO$_2$ and H$_2$O.  For H$_2$O, moist and large-scale convection were taken into account following \citet{Manabe1967}. Precipitation of H$_2$O due to coagulation was also included using a simple threshold parameterisation \citep{Emanuel1999}.

On the surface, the local albedo varied according to the composition (rocky, ocean, CO$_2$ or H$_2$O ice; see Table \ref{tab:params}). In the wet simulations, ice formation (melting) was assumed to occur when the surface temperature was lower (higher) than 273 K, and temperature changes due to the latent heat of fusion were taken into account.
In all cases, the simulations were run until collapse/glaciation occurred or steady states of thermal equilibrium were reached. The time taken to reach thermal equilibrium can be estimated from the atmospheric radiative relaxation timescale \citep{Goody1989}
\begin{equation}
\tau_r = \frac{c_p p_{s}}{\sigma g T_{e}^3},
\end{equation}
where $c_p$, $p_s$, g, $T_e$ and $\sigma$ are the specific heat capacity of the atmosphere, the mean surface pressure, the surface gravity, the atmospheric emission temperature and the Stefan-Boltzmann constant, respectively. Taking $T_e=200$ K, $c_p = 850$ J K$^{-1}$ kg$^{-1}$, $p_s = 30$ bar and $g=16.6$ m s$^{-2}$ yields $\tau_r \sim 4000$ Earth days (60 GJ581d orbits). This was close to the timescales we observed in the model by plotting time series of mean surface temperature.

We used the minimum mass for GJ581d given by \citet{Mayor2009} instead of the smaller value ($M_{min}=5.6 M_{Earth}$) proposed by \citet{Vogt2010}. We took the actual mass of GJ581d to be $M=M_{min} \slash \mbox{sin } 60^\circ = 8.2 M_{Earth}$, given that the statistically most probable value for the inclination angle is 60$^{\circ}$. Radius and gravity for rocky and ocean / ice cases (Table \ref{tab:params}) were then determined from  theoretical models \citep{Sotin2007}. In the latter case, the assumed bulk composition of the planet was 50\% H$_2$O. Model tests using $M=M_{min}$ and $M=1.6 M_{min}$ (the latter value comes from dynamical stability considerations \citep{Mayor2009}) did not reveal significant differences from our main results.

To produce emission spectra, we recorded top-of-atmosphere longitude-latitude maps of outgoing fluxes computed by the GCM over one orbit. Inclination angle of the orbit relative to the observer was assumed to be $ 60^\circ$.  An isotropic (Lambertian) distribution of specific intensities at the top of the atmosphere was assumed; comparison with a line-by-line radiative transfer code at wavelengths where the limb-darkening was most pronounced revealed that the disk-integrated flux error due to this effect was below 5\%.

\section{Results}\label{sec:results}

We performed simulations with 5, 10, 20 and 30 bar atmospheric pressure and 1:1, 1:2 and 1:10 orbit-rotation resonances for both rocky and ocean planets (see Table 1).  Eccentricity was set to zero in the simulations we present here, with the stellar flux set to the true value at $a=0.22$ and not increased to account for orbital averaging as in our 1D study \citep{Wordsworth2010b}. However, we also performed tests with $e=0.38$ and found that eccentricity was not critical to the results. Similarly, we assumed zero obliquity to assess the likelihood of atmospheric collapse in the most severe cases, but sensistivity tests showed that its influence on the climate was second order at the high pressures where the atmosphere remained stable.

CO$_2$ gas had a powerful greenhouse warming effect in our simulations because GJ581d orbits a red dwarf star. In the Solar System, Rayleigh scattering reflects the bluer incoming starlight much more effectively, and hence a planet receiving the same flux as GJ581d would be uninhabitable for any CO$_2$ pressure. In our rocky simulations, for pressures below $\sim$10 bar the atmosphere was indeed unstable, and began to condense on the dark side and / or poles of the planet (Fig. 2a). However, for denser atmospheres, we found that horizontal heat transport and greenhouse warming became effective enough to remove the threat of collapse and allow surface temperatures above the melting point of water (Fig. 1). Rotation rate affected the atmospheric stability through both the insolation (the planet has permanent day and night sides in the most extreme 1:1 resonance case) and the horizontal heat transport (faster rotating atmospheres were less efficient at transporting heat polewards).

Even in the stable simulations, some CO$_2$ usually condensed in the middle atmosphere (Fig. 3), leading to CO$_2$ ice cloud formation as occurs on present-day Mars \citep{Montmessin2007}. CO$_2$ clouds typically had a net warming effect (up to 12 K), depending on the cloud microphysical assumptions, due to the scattering of infrared radiation from the ground \citep{Forget1997}. To check the robustness of our results, we also performed tests with the (unrealistic) assumption of no CO$_2$ cloud radiative effects, and found that the atmospheres were colder, but still stable above around 10 bar.

In the ocean planet simulations, we neglected all oceanic horizontal heat transport. This assumption led us to overestimate global temperature differences and hence the probability of atmospheric collapse, in keeping with the aim of a conservative habitability estimate. We found that atmospheric H$_2$O vapour greatly increased warming, while H$_2$O cloud formation low in the atmosphere tended to cool the planet by increasing the planetary albedo (Fig. 3).  This led to a transition in the climate as total pressure increased. At 10 bar and below, cooling effects dominated and runaway glaciation occurred, followed by atmospheric collapse. From 20 bar, the positive feedback of water vapour on greenhouse warming raised mean temperatures significantly compared with the rocky case (Fig. 2b). In the intermediate region, climate stability was difficult to assess as surface temperature trends were extremely small (as low as a few K per 100 orbits in some cases). There, our results were somewhat sensitive to our microphysical assumptions; for example, lowering the precipitation threshold $l_0$ (see Table 1) resulted in optically thinner H$_2$O clouds and stable, cooler climates at lower total pressures.
Nonetheless, at higher pressures we found stable, hot climates with little global temperature variations even in the tidally locked cases, regardless of the choice of microphysical parameters.

We investigated the effect of starting with a surface covered by H$_2$O ice. For the dense atmospheres where the climate was stable, this only weakly changed the planetary albedo, and the ice melted for pressures of 20 bar or more. Hence the runaway (H$_2$O) glaciation that is possible on Earth \citep{Budyko1969} would be unlikely to occur on an ocean-planet GJ581d. Our results here contrast with those of \cite{Spiegel2009}, who used a 1D energy balance model to investigate the effects of obliquity on planetary habitability, because they neglected the dependence of planetary albedo on the total pressure in dense atmospheres.

We also investigated the possibility that an ice-covered, tidally locked GJ581d with permanent day and night sides could be locally habitable on the day side due to partial melting. However in this scenario, the day side only stayed warm when the atmosphere was thin and hence inefficient at transporting heat. As a result, the planet's dark side became cold enough for the collapse of even an N$_2$ atmosphere. An ice planet with only a thin sublimation-driven H$_2$O atmosphere could have dayside temperatures above 273 K and continual transport of H$_2$O to the dark side, but the low atmospheric pressure would preclude liquid water except in extremely limited sub-surface regions. Clearly, dense, stable atmospheres offer better prospects for habitability.

\section{Discussion}\label{sec:conc}

Are CO$_2$ partial pressures of over 10 bar a realistic possibility for GJ581d? Given the planet's assumed gravity, this corresponds to a CO$_2$ column of only 4 to 6 bars on Earth. Even though Venus and Earth are smaller,  their total CO$_2$ inventories scaled by planetary mass are around 10-100 times greater than this. On Venus, it is thought that a large fraction of this inventory is in the atmosphere \citep{Bullock1996}, while on Earth, the atmospheric partial pressure is regulated over geological timescales by the carbonate-silicate cycle \citep{Walker1981} via plate tectonics. The geophysics of GJ581d is unknown, but if a similar mechanism were present there, its atmospheric CO$_2$ would stabilize above the level needed to maintain a liquid water cycle by negative feedback.

CO$_2$-H$_2$O-(buffer gas) atmospheres are clearly only a subset of all the possible scenarios for GJ581d. If small quantities of additional greenhouse gases such as CH$_4$ or SO$_2$ were present, they would increase warming further. However, if these gases were chemically unstable, they would need to be continually emitted by e.g., volcanic or biological sources to have a long-term effect on the climate. More drastically, the planet could  have a thick H$_2$-He envelope like Uranus or Neptune, or its atmosphere could have been removed entirely by increased stellar activity early in the life of the host star. Both scenarios would exclude surface liquid water (the former through an excess of surface pressure). GJ581d's large mass means that it can retain an atmosphere more easily than Earth or Venus, but the increased XUV and ion fluxes from M-class stars early in their history means that removal processes can be much more efficient than in the Solar System \citep{Tanigawa2007,Tian2009}. The escape of atmospheric hydrogen depends sensitively on both XUV flux levels and radiative cooling by H$_3^+$ ions, both of which  are poorly constrained for GJ581d \citep{Selsis2007,Koskinen2007}.
Moreover, stripping by the stellar wind, which does not discriminate between light and heavy atoms, could have removed large amounts of any gas (including CO$_2$) \citep{Lammer2007}. Detailed modelling of atmospheric escape may be able to constrain the effects of these processes on the atmosphere further.

Unlike the majority of the Kepler planetary candidates, Gliese 581d is relatively close to Earth, so in the future it will be possible to establish which scenario applies to it through direct spectroscopic observations. To determine the sensitivity required for this, we used the outgoing longwave radiation (OLR) from our simulations to produce synthetic emission spectra for the habitable climates just discussed, along with those for a planet with no atmosphere (Fig. 4).  We did not calculate emission spectra for the H$_2$-He dominated case, but equilibrium chemistry calculations\footnote{These were performed using a code developed by R. Bounaceur (LRGP, CNRS, Nancy, France), based on an algorithm to minimize the free energy of a given mixture of gases \citep{Reynolds1986}.} allowed us to establish that for an atmosphere where H$_2$ is the dominant species by mass, the abundance of CO$_2$ should be below the limit of detectability. For a rocky / icy planet where volatiles like CO$_2$ have collapsed on the surface and the atmosphere is thin or non-existent, the flux variations over one orbit will be large (grey region in Fig. \ref{fig:spectra}). Observations of CO$_2$ and H$_2$O absorption bands with low phase variations (red / blue regions in Fig. \ref{fig:spectra}) would hence be a strong indicator of the kind of stable, dense habitable atmospheres discussed in this paper. From Fig. \ref{fig:spectra}, the flux sensitivity necessary for this in the infrared will be of order 10$^{-21}$ W m$^{-2}$ $\mu$m$^{-1}$, corresponding to a planet / star ratio lower than $10^{-6}$ for wavelengths $<$ 20~$\mu$m. This is beyond the capabilities of current space and ground-based observatories, but it will become possible with future improvements in instrumental technology.

\acknowledgments

F.S. acknowledges support from the European Research Council (Starting Grant 209622: E$_3$ARTHs).

\bibliographystyle{plainnat}

\begin{table}[h]
\centering
\caption{Standard parameters used in the climate simulations.}
\begin{tabular}{lll}
\hline \hline
Stellar luminosity & $L$ [$L_{Sun}$] & 0.0135\\
Orbital semi-major axis & $a$ [AU] & 0.22 \\
Orbital eccentricity & $e$ & 0.0 \\
Obliquity & $\phi$ & 0.0 \\
Tidal resonance & $n$ &  1, 2, 10 \\
Initial atmospheric pressure & $p_{s}$ [bars] &  5-30 \\
Radius (rocky) & $r$ [r$_{Earth}$] & 1.8 \\
Radius (ocean) & $r$ [r$_{Earth}$] & 2.3 \\
Surface gravity (rocky) & $g$ [m s$^{-2}$] & 25.0 \\
Surface gravity (ocean) & $g$ [m s$^{-2}$] & 16.6 \\
Surface albedo (rocky) & $A_s$ & 0.2 \\
Surface albedo (ocean) & $A_s$ & 0.07 \\
Surface albedo (ice) & $A_s$ & 0.6 \\
Surface roughness coefficient & $z_0$ [m] & $1\times10^{-2}$ \\
Precipitation threshold & $l_0$ [kg kg$^{-1}$] & 0.001 \\
Number of cloud condensation nuclei & $N_c$ [kg$^{-1}$] & $1\times10^5$ \\
\hline \hline
\end{tabular}\label{tab:params}
\end{table}

\begin{figure}[h]
	\begin{center}
		{\includegraphics[width=2.8in]{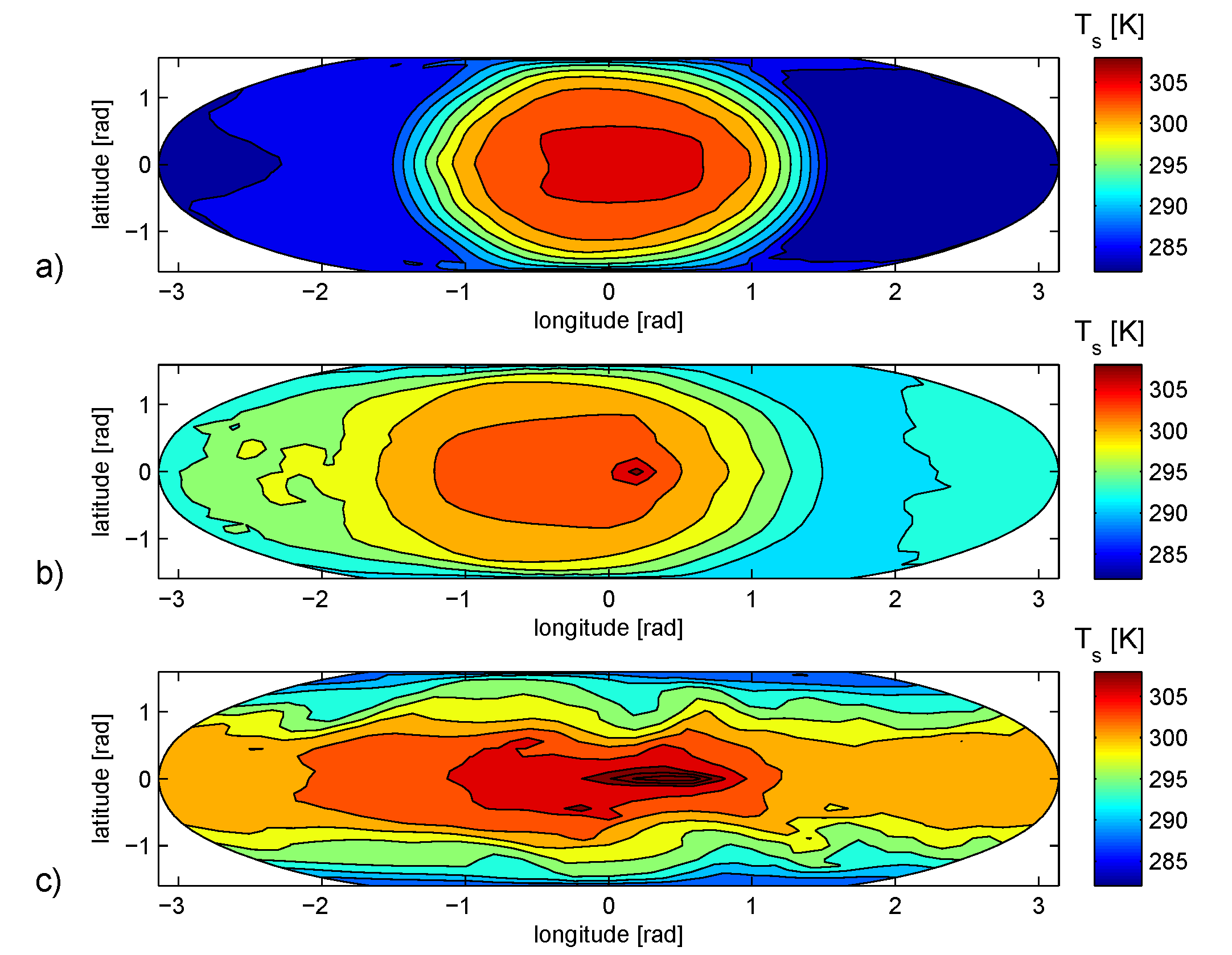}}
	\end{center}
	\caption{Surface temperature snapshots after 60 orbits integration time for rocky planet simulations with a) 1:1, b) 1:2 and c) 1:10 tidal resonance and a 20-bar CO$_2$ atmosphere.}
	\label{fig:snapshots}
\end{figure}

\begin{figure}[h]
	\begin{center}
		{\includegraphics[width=2.8in]{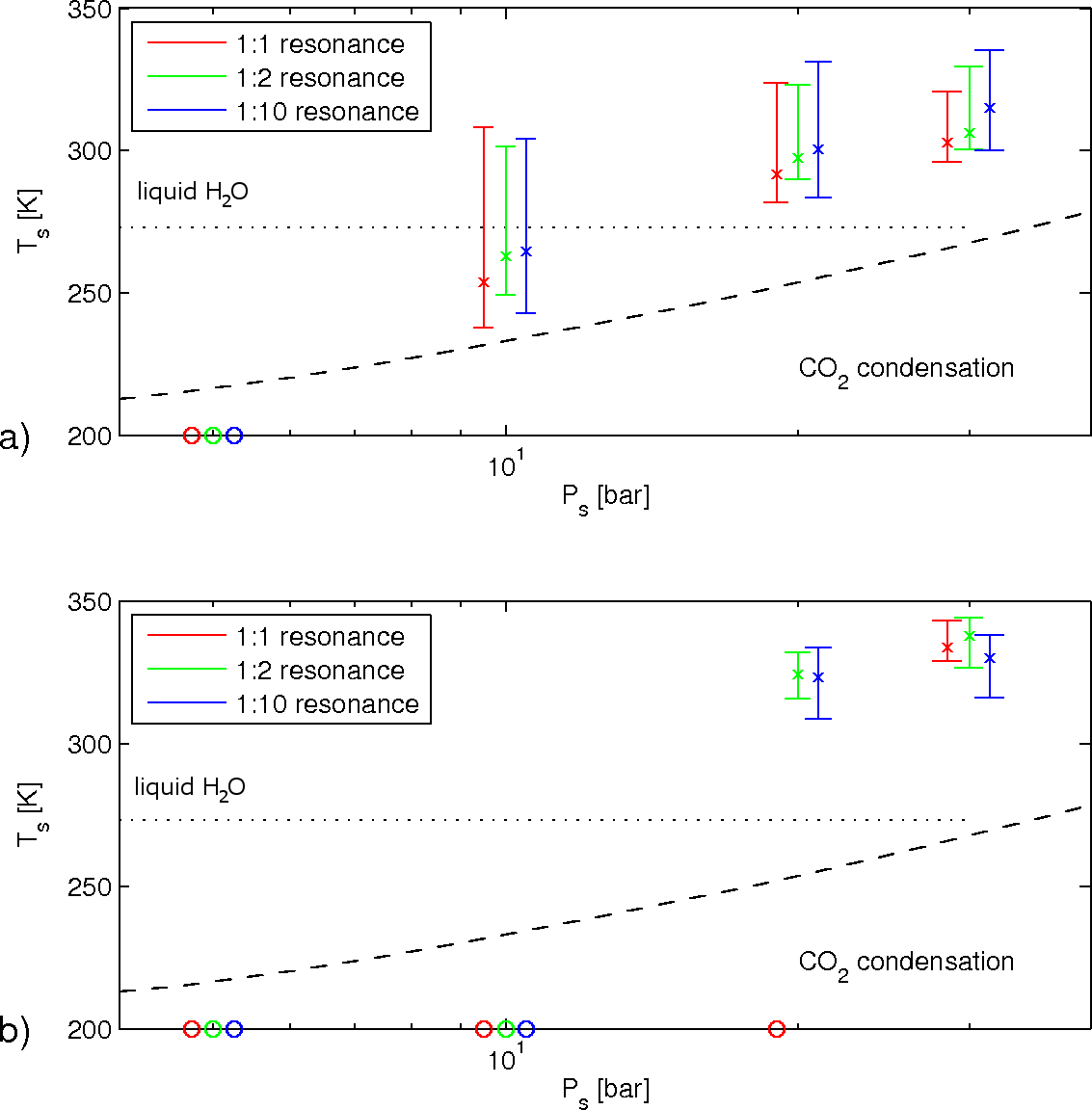}}
	\end{center}
	\caption{Surface temperature vs. pressure for a) rocky and b) ocean planet simulations. Red, green and blue markers represent tidal resonances of 1:1, 1:2 and 1:10. Points are separated for clarity; simulations were performed at 5, 10, 20 and 30 bar for each all cases. Crosses and error bars show mean and maximum / minimum temperatures (sampled over one orbit and across the planet's surface) in stable cases, while circles indicate simulations where the atmosphere began to collapse or runaway glaciation occurred.}
	\label{fig:TsPs}
\end{figure}

\begin{figure}[h]
	\begin{center}
		{\includegraphics[width=2.8in]{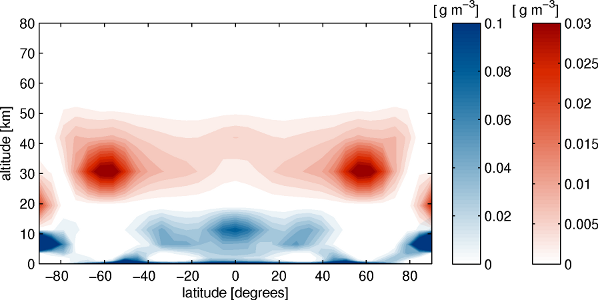}}
	\end{center}
	\caption{Latitude-altitude map of H$_2$O (blue) and CO$_2$ (red) cloud coverage for a 20-bar ocean simulation with 1:2 resonance, averaged in longitude and over one orbit. While the cloud deck altitudes were similar for all simulations, the latitudinal distribution depended strongly on the atmospheric dynamics and hence on the rotation rate and total pressure.}
	\label{fig:clouds}
\end{figure}

\begin{figure}[h]
	\begin{center}
		{\includegraphics[width=2.8in]{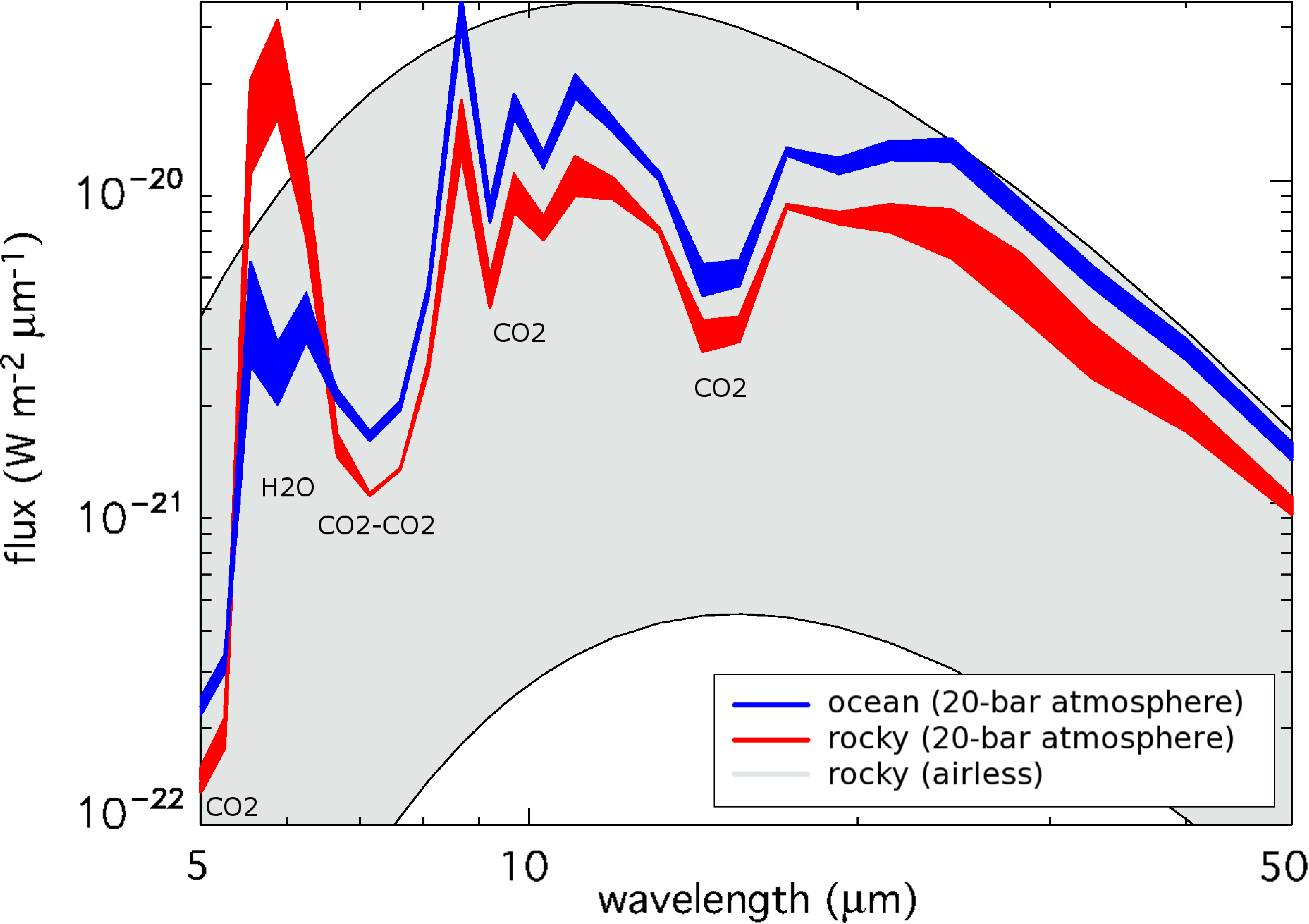}}
	\end{center}
	\caption{Infrared emission spectra for an ocean / rocky GJ581d with a 20-bar atmosphere (blue / red) and a rocky GJ581d with no atmosphere (grey). Thickness of the lines corresponds to the maximum / minimum difference in flux over one orbit, for the most probable observation angle of 60$^\circ$. The maximum (minimum) surface temperature in the airless case is 271 K (37 K). For the cases with atmospheres, emission of the ocean planet is higher at most wavelengths due to the increased planetary radius. The major absorption features are labelled by molecule / process (CO$_2$-CO$_2$ corresponds to CO$_2$ collision-induced absorption).}
	\label{fig:spectra}
\end{figure}

\end{document}